\documentclass[twocolumn, pra, showpacs, aps, superscriptaddress]{revtex4-1}

\usepackage{graphicx}
\usepackage{ae,aecompl}
\usepackage{bm}
\usepackage{textcomp}
\usepackage{xcolor}

\usepackage{wasysym}
\usepackage{hyperref}

\newcommand{\2}{\downarrow}
\newcommand{\1}{\uparrow}
\newcommand{\upket}{|\!\!\1\rangle}
\newcommand{\downket}{|\!\!\2\rangle}

\newcommand{\phdag}{^{\vphantom{dag}}}
\newcommand{\phb}{_{\1\vphantom{b}}}

\begin{document}


\date{\today}

\title{Twist-and-Turn Spin Squeezing in Bose-Einstein Condensates}

\author{W. Muessel}
\email{TnT@matterwave.de}
\author{H. Strobel}
\author{D. Linnemann}

\affiliation{Kirchhoff-Institut f\"ur Physik, Universit\"at Heidelberg, Im Neuenheimer Feld 227, 69120 Heidelberg, Germany.}

\author{T. Zibold}
\altaffiliation{Present address: Laboratoire Kastler Brossel, Coll{\`e}ge de France, 11 place Marcelin Berthelot, 75005 Paris, France}
\affiliation{Kirchhoff-Institut f\"ur Physik, Universit\"at Heidelberg, Im Neuenheimer Feld 227, 69120 Heidelberg, Germany.}

\author{B. Juli{\'a}-D{\'i}az}
\affiliation{Departament d'Estructura i Constituents de la Mat\`{e}ria,
Universitat de Barcelona, Spain}
\affiliation{ICFO-Institut de Ci\`encies Fot\`oniques, Parc Mediterrani 
de la Tecnologia, 08860 Barcelona, Spain}
%
%

\author{M.~K. Oberthaler}
\affiliation{Kirchhoff-Institut f\"ur Physik, Universit\"at Heidelberg, Im Neuenheimer Feld 227, 69120 Heidelberg, Germany.}


\begin{abstract}

We experimentally demonstrate an alternative method for the dynamic generation of atomic spin squeezing, building on the interplay between linear coupling and nonlinear phase evolution.
Since the resulting quantum dynamics can be seen as rotation and shear on the generalized Bloch sphere, we call this scheme twist-and-turn (TnT). 
This is closely connected to an underlying instability in the classical limit of this system.
The short-time evolution of the quantum state is governed by a fast  initial spreading of the quantum uncertainty in one direction, accompanied by squeezing in the orthogonal axis. We find an optimal value of $\xi_{\text{S}}^2=-7.1(3)$\,dB in a single BEC and scalability of the squeezing to more than $10^4$ particles with $\xi_{\text{S}}^2=-2.8(4)$\,dB.
\end{abstract}
\maketitle 

\textit{Introduction.} The efficient generation of highly entangled states is among the biggest challenges in quantum technologies, as they allow approaching the ultimate quantum limits like the elusive Heisenberg limit in metrology \cite{GiovannettiSCIENCE2004}. 
Within the last decade, technological development has enabled a number of schemes that reliably produce entangled  many-particle quantum states, ranging from spin squeezed to Dicke states for neutral atoms \cite{EsteveNATURE2008, AppelPNAS2009, GrossNATURE2010, RiedelNATURE2010, LerouxPRL2010-CS, SchleierSmithPRL2010, ChenPRL2011, SewellPRL2012, HamleyNATPHYS2012, BerradaNATCOMM2013, HaasSCIENCE2014,LueckePRL2014,McConnellNATURE2015}.\\
The Lipkin-Meshkov-Glick Hamiltonian, originally developed in nuclear physics \cite{LipkinNUCLPHYS1965}, captures the dynamics of interacting particles in two modes. It allows the generation of a rich variety of spin squeezed \cite{DiazPRA2012} and  highly entangled non-Gaussian states \cite{MicheliPRA2003}. In ultracold gases, this Hamiltonian can be implemented using linear Rabi coupling and atomic interactions between two internal states. With that system, entangled non-Gaussian states have been created \cite{StrobelSCIENCE2014}. In this work, we detail the dynamic generation of spin squeezed states.\\
The addition of linear coupling to the standard one-axis-twisting scheme \cite{KitagawaPRA1993} leads to an exponentially increasing quantum uncertainty, implying a corresponding squeezing along the orthogonal axis on a fast time scale.
This results from the interplay of the twist due to the interaction and the rotation (turn) caused by the linear coupling. 
Semiclassically, this behavior can be understood as the dynamics  of a classical phase space volume around an unstable fixed point. The corresponding emergence of correlations has also been studied in a three-mode scenario \cite{HamleyNATPHYS2012, GervingNATCOMMUN2012}.  
\begin{figure}[b!]
\includegraphics[width = 86mm]{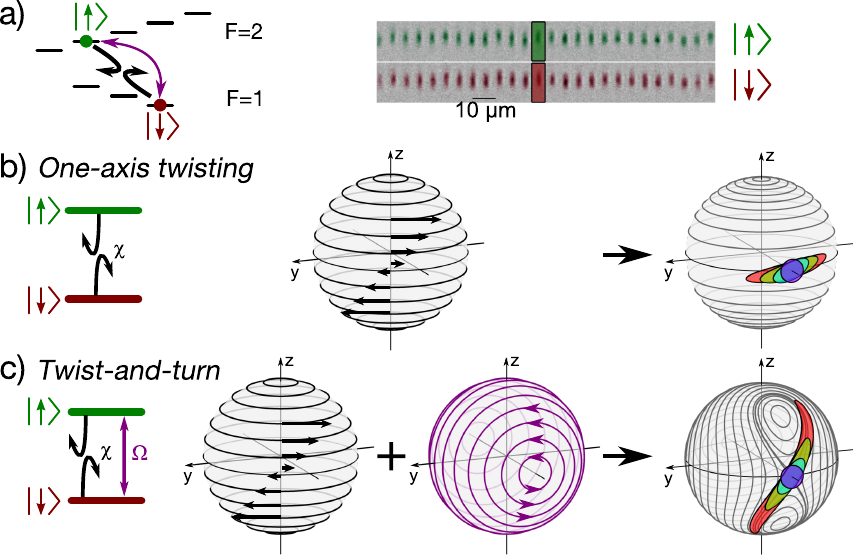}
\caption{(color online). {\bf Experimental system and classical phase space.} (a) We simultaneously generate around 30 BECs, each containing $N =200-600$ atoms,  using an optical lattice potential. We employ two internal states $\downket = |F=1, m_F = +1\rangle$ and $\upket = |2, -1\rangle$ of  $ ^{87}$Rb. The populations of the two states are read out by absorption imaging after Stern-Gerlach separation. b) The nonlinearity $\chi$ induces an angular velocity which depends on the population imbalance  and twists the quantum uncertainty. The squeezing dynamics of this one-axis twisting scheme in a single BEC is shown on a generalized Bloch sphere (right panel). c) The interplay of the nonlinearity and linear coupling (rotation around the $x$-axis with angular velocity $\Omega$) creates a twist-and-turn scenario. The resulting dynamics features initial squeezing and emergence of non-Gaussianity at later times.}\label{Fig1}
\end{figure}

\textit{Experimental system.} In our experiment, the Lipkin-Meshkov-Glick Hamiltonian is realized by employing two internal states of a $^{87}$Rb Bose-Einstein condensate \cite{ZiboldPRL2010}, the $\downket= |F=1, m_F = +1\rangle$ and $\upket = |2, -1\rangle$ hyperfine states of the electronic ground state.  We simultaneously prepare up to 30 independent condensates, each containing $N=200-600$ atoms, in an optical standing wave potential \cite{StrobelSCIENCE2014}. The large trapping frequencies ensure that external dynamics is frozen out. Additionally, the array of condensates yields many independent realizations and enables scalability to large atom numbers  \cite{MuesselPRL2014}.\\
Linear coupling between the two internal states is achieved using a two-photon microwave ($\approx 6.8$\,GHz) and radio frequency ($\approx 6.3$\,MHz) transition. The phase of the coupling can be nonadiabatically adjusted by switching the phase  of the radio frequency radiation, which is created using an arbitrary waveform generator.  In addition, the coupling power can be altered by changing the amplitude of the radio frequency signal and attenuating the microwave by use of a fixed attenuator on an RF switch with two ports. The atomic nonlinearity is enhanced  in the vicinity of an interspecies Feshbach resonance between  $\upket$ and  $\downket$ at a magnetic bias field of 9.12\,G. After each experimental cycle, state-dependent detection is implemented using absorption imaging after Stern-Gerlach separation of the two components \cite{Muessel2013}  (see Fig.\,\ref{Fig1}a). The imaging is performed at low magnetic fields ($\sim 1$\,G) after a ramp-down of the bias field in 300\,ms. In order to inhibit spin-relaxation loss of the $\upket$ state during the ramp-down, we transfer its population to $|1,-1\rangle$ via a microwave $\pi$ pulse.

\textit{Theoretical description.} 
In a quantum mechanical description, the system of $N$ indistinguishable two-level bosons can be treated as a pseudospin ($J = N/2$) and displayed on a generalized Bloch sphere. Using the respective creation and annihilation operators of the two modes, the $z$ component of the pseudospin $\hat{J}_z = \frac{1}{2}(\hat{a}\phb^\dag\hat{a}_\1\phdag-\hat{a}_\2^\dag\hat{a}_\2\phdag) =  (\hat{N}_{\uparrow}-\hat{N}_{\downarrow})/2$ is defined by the population difference between the  two levels, which can be directly detected in the experiment. The orthogonal components (coherences) are given by
$\hat{J}_x = \frac{1}{2}(\hat{a}\phb^\dag\hat{a}_\2\phdag+\hat{a}_\2^\dag\hat{a}_\1\phdag)$  and 
$\hat{J}_y = \frac{1}{2\text{i}}(\hat{a}\phb^\dag\hat{a}_\2\phdag-\hat{a}_\2^\dag\hat{a}_\1\phdag)$, fulfilling the angular momentum commutation relation $[\hat{J}_j, \hat{J}_k] = \text{i} \epsilon_{jkl}\hat{J}_l$.\\
In this pseudospin picture, our experimental system can be described by the Hamiltonian
\begin{equation}
\hat{\mathcal{H}} = \chi \hat{J}_z ^2 - \Omega \hat{J}_x + \delta \hat{J}_z,
\label{EqHamiltonian}
\end{equation}
which is a special case of the Lipkin-Meshkov-Glick Hamiltonian \cite{LipkinNUCLPHYS1965}. The quantum dynamics of this system is governed by the relative strength of the  three parameters: the nonlinearity $\chi$ arising from the interparticle interaction, coupling strength $\Omega$ given by the microwave and radio frequency radiation, and the detuning  $\delta$ resulting from the mismatch of the coupling and the atomic frequency, the AC Zeeman shift as well as the particle number dependent mean field shifts.\\
The resulting dynamics can be understood from the fact that angular momentum operators are the generators of rotations. Thus, $(\chi \hat{J}_z) \hat{J}_z$ leads to a rotation around the $z$ direction whose angular velocity $\chi \hat{J}_z$ depends on the population difference $\hat{J}_z$. This can be interpreted as a twist (Fig.\,\ref{Fig1}b). The one-axis twisting scheme exploits this nonlinear twist exclusively.\\
The second and third term in the Hamiltonian describe linear rotations around the $x$- and the $z$-axis, respectively. In the regime $\Lambda = |N\chi/\Omega|> 1$,  the relative sign between linear coupling $\Omega$ and nonlinearity $\chi$ can be chosen such that the linear rotation  leads to a speed-up of the shear of the quantum state. This is achieved when the linear coupling transfers the enlarged spread in phase ($\hat{J}_x$) into an increased spread in particle number  difference ($\hat{J}_z$), which implies faster twisting.

\begin{figure}[t]
\includegraphics[width = 86mm]{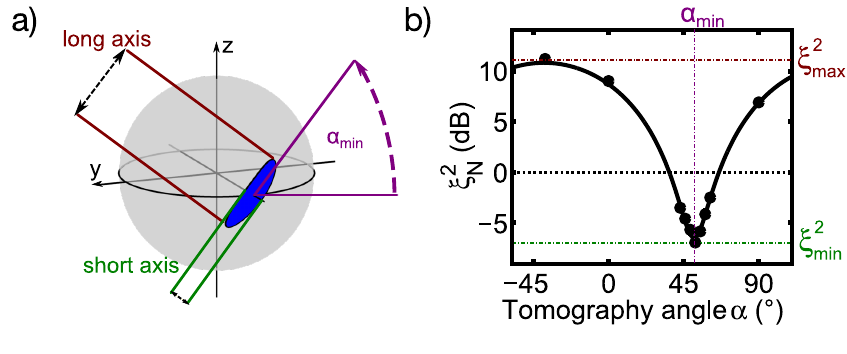}
\caption{(color online). \textbf{Squeezing analysis.} a) The generated states are tomographically analyzed by rotation around the $x$-axis and detection of the population difference ${N}_{\uparrow}-{N}_{\downarrow}$. The uncertainty of a squeezed state is characterized from the fluctuations of repeated experiments using three parameters: The extension along the elongated direction (long axis, $\xi^2_{\text{max}}$), the size of the minimal fluctuations (short axis, $\xi^2_{\text{min}}$), and the optimal tomography rotation angle $\alpha_{\text{min}}$. b) An exemplary tomography result after 15\,ms of TnT shows a strong modulation of the observed variances and suppression below the classical limit (dotted line).}
\label{Fig2}
\end{figure}
\begin{figure}[t]
\includegraphics[width = 86mm]{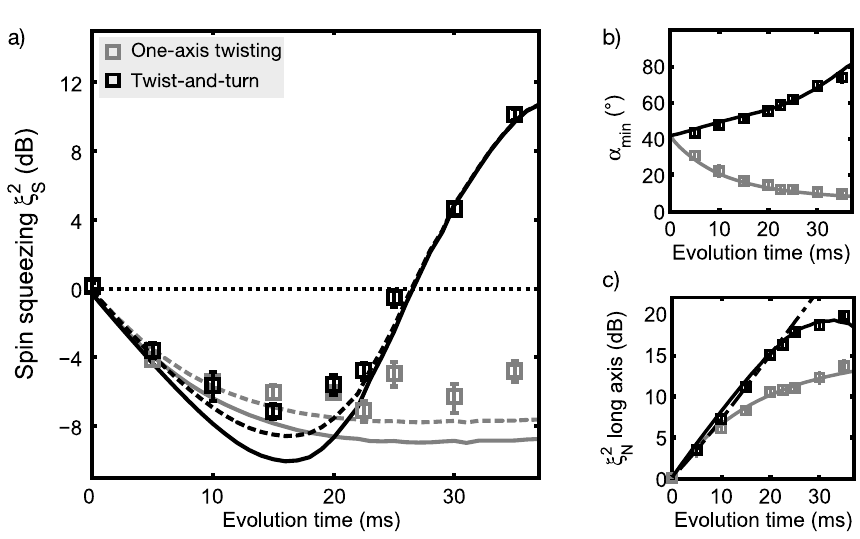}
\caption{\textbf{Spin squeezing dynamics.} (a) Spin squeezing for different evolution times obtained from one-axis twisting (gray squares) and TnT squeezing (black squares), which vanishes as the state becomes non-Gaussian. The minimal obtained value for the TnT scenario is $-7.1(3)$\,dB after 15\,ms evolution time. The experimental results are in agreement with a numerical Monte Carlo wave function analysis for the experimental parameters, which includes the effects of loss (solid lines). Additionally including known sources of noise during readout yields better quantitative agreement (dashed lines). The dotted line depicts the classical limit. (b) In contrast to the one-axis twisting scheme (gray), the optimal tomography angle for the TnT scheme (black) increases with time. (c) The fluctuations along the long axis increase exponentially for the first 25\,ms in the TnT scheme (black squares, dash-dotted line: exponential fit), indicating the underlying classical instability. Error bars are statistical 1~s.d. confidence intervals.}
\label{Fig3}
\end{figure}
\textit{Classical phase space picture.}
 Further insight can be gained by analyzing the corresponding classical description, which is valid in the limit $N\to \infty$. The corresponding classical Hamiltonian is
\begin{equation}
\mathcal{H}_{\text{class}} = \frac{N^2\chi}{4}z^2 - \frac{N\Omega}{2}\sqrt{1-z^2}\cos{\phi}+\frac{N\delta}{2}z,
\label{EqHamiltonMeanFieldLMG}
\end{equation}
with the imbalance $z = ({N}_{\uparrow}-{N}_{\downarrow})/N$ and the phase $\phi = \text{arctan}(\langle \hat{J}_y\rangle /\langle \hat{J}_x\rangle)$.
 In the case of sole twisting, the phase space portrait features two stable fixed points at the north and south pole of the generalized Bloch sphere. Exemplary classical trajectories are visualized by solid lines in Fig.\,\ref{Fig1}b.  The addition of linear coupling leads in the case of dominating interaction, i.e. $\Lambda >1$, to two additional fixed points on the equator of the Bloch sphere, one of which is stable and one unstable \cite{SmerziPRL1997, RaghavanPRA1999,ShchesnovichPRA2008,ZiboldPRL2010}. The TnT scenario exploits this unstable fixed point and is experimentally realized for $\Lambda\approx 1.5$ and ${\delta} \approx 0$ (Fig.\,\ref{Fig1}c).  The instability leads to rapid spreading of the quantum state along the separatrix which divides the classical phase space into regions of macroscopically different temporal behavior. Spin squeezing is generated during the early dynamics \cite{DiazPRA2012}. At later times, the bending around the stable fixed points leads to the appearance of non-Gaussianity, and squeezing vanishes \cite{MicheliPRA2003, StrobelSCIENCE2014}.\\
In contrast, the quantum states created by the one-axis-twisting scheme \cite{KitagawaPRA1993, SorensenNATURE2001,GrossNATURE2010, RiedelNATURE2010, LerouxPRL2010-CS} remain Gaussian on much longer timescales.

\textit{State preparation.} 
In the experiment, after initial preparation of all atoms in $\downket$, a coherent superposition  between $\upket$ and $\downket$ is produced by applying a  $\pi/2$ pulse of the linear two-photon coupling. Subsequently, microwave and radio frequency are attenuated to reach the regime of $\Lambda \approx 1.5$ and the phase of the Rabi coupling is adjusted by $3\pi/2$, which changes the rotation axis from the  $y$ to the negative $x$-axis. Additional phase shifts due to the microwave attenuator are compensated by shifting the phase of the radio frequency accordingly, and  the frequency  is  adjusted, taking into account the change in AC Zeeman shift caused by the altered power of the coupling radiation.\\
The influence of technical detuning fluctuations, caused by variations of the magnetic bias field of $\approx 45$\,\textmu G over several days, is reduced by applying a spin-echo pulse ($\pi$ rotation around the $x$ direction) at half the evolution time. This also reduces the sensitivity to coupling phase errors. By omitting the linear coupling during the evolution time, the same sequence is used for a direct comparison to the one-axis twisting scenario.


\textit{State analysis.} 
To investigate the states generated by the twist-and-turn scheme, we perform a tomographic readout. This is achieved by rotating the state around the $x$-direction for various angles and analyzing the fluctuations of the particle number difference for repeated measurements (Fig.\,\ref{Fig2}a). The fluctuations are quantified by the number squeezing parameter $\xi_{\text{N}}^2 = \text{Var}(N_\1-N_\2)/\text{Var}_{\text{class}}(N)$, normalized to the binomial variance of the corresponding coherent spin state $\text{Var}_{\text{class}}(N) = 4p(1-p)N$ with $p=\langle N_\1 \rangle/N$.
 One representative experimental result is shown in Fig.\,\ref{Fig2}b, from which we extract the maximal observed variance as well as the minimal fluctuations $\xi_{\text{min}}^2 = \text{min}(\xi_{\text{N}}^2)$ and the corresponding rotation angle $\alpha_{\text{min}}$.
 
\textit{Experimental results.}
 The generation of squeezing with the TnT scheme is quantified in Fig.\,\ref{Fig3} by the resulting spin squeezing parameter   $\xi_{\text{S}}^2 = \xi_{\text{min}}^2/ \langle\cos\varphi\rangle ^2$, taking into account the reduced mean spin length $N\langle\cos\varphi\rangle /2$ due to the extension of the state along the long axis. 
We infer $\langle\cos\varphi\rangle$ by applying a rotation with an angle $\pi/2 -\alpha_{\text{min}}$ and detecting the distribution of particle number differences. Since the population imbalance after the rotation is $z = \sin{\varphi}$, we can directly access the expectation value $\langle\cos\varphi\rangle$\cite{GrossNATURE2010}.\\  
We experimentally find strong initial spin squeezing, reaching a minimal value of $\xi_{\text{S}}^2 = -7.1(3)$\,dB after 15\,ms evolution time (Fig.\,\ref{Fig3}a). The precisely characterized photon shot noise of the absorption images (standard deviation $\sigma_{\text{det}} \approx 4$\,atoms for each component) has been subtracted for all given values. Small fringe noise contributions remain \cite{Muessel2013}. Our results confirm the generation of entanglement during the early evolution.\\
 After this minimal value,  squeezing is quickly lost. This does not imply the loss of entanglement, as spin squeezing only captures the variance properties of the state and thus does not fully characterize non-Gaussian states. In this regime, entanglement can be shown by extraction of the Fisher information \cite{StrobelSCIENCE2014}.\\
The solid lines in Fig.\,\ref{Fig3}a represent the results of a Monte Carlo wave function (MCWF) simulation discussed in detail in the Appendix, and qualitatively agrees with the experimental data.
This simulation includes particle losses and the resulting change of the parameters $\chi(N)$ and $\delta(N)$ which depend on the atomic density, as well as the technical detuning fluctuations. The dashed line includes additional known sources of noise during the detection process, such as the residual losses due to background collisions during the ramp-down of the magnetic field, which corresponds to a loss of $\approx 8$ atoms.\\
We also compare this novel scheme with the well-established one-axis twisting scheme. While the corresponding experimentally obtained value of the best squeezing (gray squares in Fig.\,\ref{Fig3}a) is comparable, the transition to non-Gaussian states happens at much larger time scales, which are longer than the largest investigated evolution time.\\ 
The experimentally extracted angle of minimal fluctuations $\alpha_{\text{min}}$ and the size of the long axis (see Fig.\,\ref{Fig3}b and c) are much less susceptible to loss than the minimal squeezing, yielding very good agreement with our MCWF simulations. The underlying classical instability in the TnT scheme leads to an exponential growth of quantum mechanical uncertainty, which is directly observed by analyzing the increase of fluctuations of the  long axis shown in Fig.\,\ref{Fig3}c. This is in contrast to one-axis twisting, where the deviation  from the initial exponential growth occurs much earlier \cite{DiazPRA2012}, and confirms that the exponentially fast initial squeezing is prolonged in the TnT scheme. 


\begin{figure}[t]
\includegraphics[width = 86 mm]{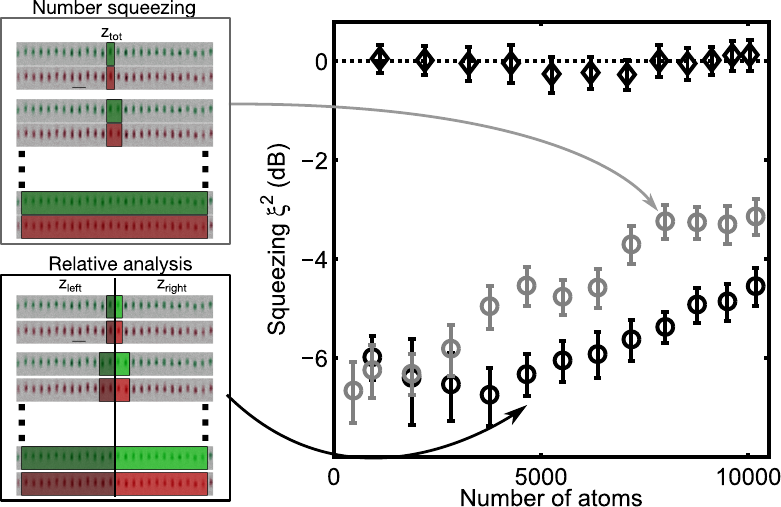}
\caption{(color online). {\bf Scalability to large atom numbers.} The best obtained number squeezing after 15\,ms of TnT evolution can be scaled to large particle number by summing up the atom numbers of adjacent sites (upper left panel), yielding $\xi_{\text{N}}^2 = -3.1(4)$\,dB for 10\,200 particles (gray circles). A differential analysis of the fluctuations between two different parts of the array (lower  left panel) is more robust against technical fluctuations and improves this value to $-4.5(4)$\,dB. The remaining decrease of the observed fluctuation suppression is  caused by the atom number inhomogeneity of the lattice, which leads to different optimal rotation angles. As a reference, the scaling for a coherent spin state is indicated as black diamonds and is in good agreement with the classical limit (dotted line). Error bars are 1 s.d. confidence intervals from a resampling analysis \cite{MillerBIOMETRIKA1974}.}
\label{Fig4}
\end{figure}

\begin{figure}[b]
\includegraphics{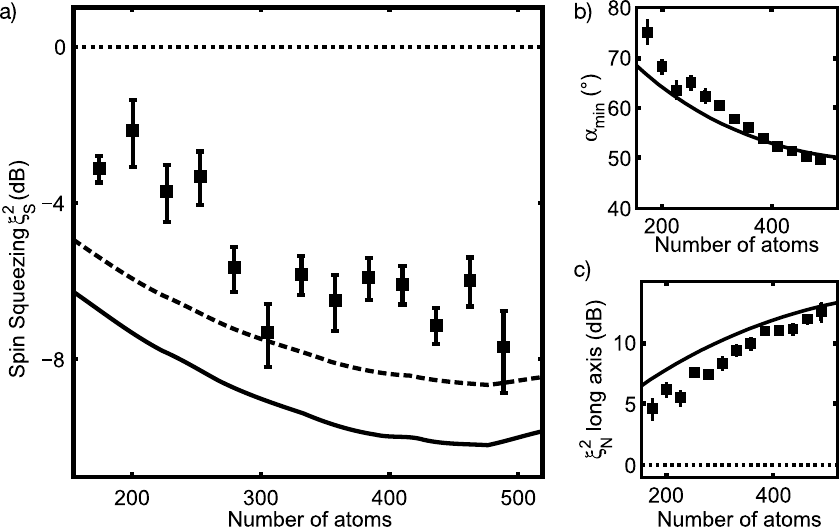}
\caption[width = 86 mm]{{\bf Atom number dependence  in the single BECs.} Due to the criticality of the TnT scheme and the atom number dependence of $\chi$ and $\delta$, the characteristics of final states after a fixed evolution time change with atom number. This is shown for an evolution time of $15$\,ms and three key parameters of the spin squeezed states: a) The spin squeezing parameter $\xi_{\text{S}}^2$, b) the optimal rotation angle $\alpha_{\text{min}}$ and c) the extension of the long axis. The results are reproduced by a MCWF simulation including the experimental parameter dependences on atom number (solid line). The dashed line indicates the results including additional noise during the readout sequence, the dotted lines are the corresponding classical limits. Error bars are statistical 1 s.d. confidence intervals.}
\label{Fig5}
\end{figure}

\textit{Upscaling of  TnT to large atom numbers.}
The obtainable phase precision in an interferometric sequence is limited by $\Delta \phi = \xi_{\text{S}}/\sqrt{N}$ for a spin squeezed state with  $N$ particles and spin squeezing parameter $\xi_{\text{S}}$. Therefore, it is one of the main challenges to generate squeezing also for large atom numbers. It has been shown that the use of many mesoscopic condensates allows the upscaling  of squeezing to large numbers \cite{MuesselPRL2014} by adding up the populations of the individual condensates. \\
To analyze the number squeezing parameter for different ensemble sizes, we sum the atom numbers  of different lattice sites ${N}_{\1{\text{tot}}} = \sum_i N_{\1i}$ and ${N}_{\2{\text{tot}}} = \sum_i N_{\2i}$ (upper left panel of Fig.\,\ref{Fig4}) and calculate $\xi_{\text{N}}^2$ in analogy to the evaluation for the single sites. For an evolution time of 15\,ms and the optimal angle $\alpha = 52$\textdegree, this analysis yields a noise suppression $\xi_{\text{N}}^2 = -3.1(4)$\,dB even for the full ensemble of 10\,200 particles (gray circles in Fig.\,\ref{Fig4}). As the classical reference, the corresponding scaling for a coherent spin state is in agreement with the shot-noise limit for all atom numbers (diamonds).\\
The loss of squeezing for large atom numbers is caused both by atom number inhomogeneities over the array and technical noise sources, which are dominated by fluctuations of the magnetic bias field. The influence of the latter can be minimized by employing a differential analysis, in which the array is divided in two parts and the relative fluctuations of the two population imbalances $z_{\text{left}}$ and $z_{\text{right}}$  are analyzed. This is robust against the technical fluctuations, as these are suppressed for the difference $\delta z = z_{\text{left}}- z_{\text{right}}$ \cite{MuesselPRL2014}. The corresponding relative squeezing parameter $\xi_{\text{Rel}}^2 = \text{Var}(\delta z)/\text{Var}(\delta z)_{\text{class}}$ quantifies the noise suppression relative to the classical limit $\text{Var}(\delta z)_{\text{class}}$. With this analysis, we find $\xi_{\text{Rel}}^2 = -4.5(4)$\,dB for the full sample. This is on a comparable level with the value of  $\xi_{\text{Rel}}^2 = -5.3(5)$\,dB that can be obtained using one-axis twisting \cite{MuesselPRL2014}. \\
The remaining decrease of squeezing can be attributed to the atom number dependence of the TnT squeezing scheme (Fig.\,\ref{Fig5}), for which the degree of single-site squeezing improves as the particle number increases. This results from the particle number dependence of the detuning since the intra-species scattering lengths of the two components are slightly different.  The numerical solution of the stationary Gross-Pitaevskii equation for our experimental situation reveals to good approximation the scalings  $\delta\propto\sqrt{N}$ and  $\chi\propto1/\sqrt{N}$. Both effects are included in the MCWF simulation (solid lines). Since we fix the angle of rotation to $\alpha = 52$\textdegree, the strong dependence $\alpha_{\text{min}}$ on atom number (see Fig.\,\ref{Fig5}b) mainly limits the scalability using mesoscopic samples with varying atom number. Note that this atom number dependence of the final state is stronger compared to the one-axis-twisting scheme, which leads to the slightly lower value for the relative squeezing of the full ensemble.\\
For the sum of several independent condensates, the two-mode approximation is not valid, and the mean spin length has to be extracted from the visibility $\mathcal{V}$ of Ramsey fringes. As we observe $\mathcal{V} = 94.2$\% for the full ensemble, the whole resource can be directly exploited for quantum enhanced measurements, either in a DC or a gradiometric scheme. The corresponding spin squeezing parameters \cite{WinelandPRA1994} are $\xi_{\text{S}}^2 = -2.8$\,dB for the direct analysis, and  $\xi_{\text{Sdiff}}^2 = -4.0$\,dB for the differential case, showing the applicability for quantum-enhanced metrology.


\textit{Conclusion.} 
We have shown that the twist-and-turn scheme can efficiently generate spin squeezing on short experimental time scales and is thus favorable for squeezing in lossy environments. These spin squeezed states can be directly employed in quantum-enhanced measurement schemes, which have already been demonstrated for clocks and magnetometry \cite{LouchetChauvetNJP2010, LerouxPRL2010, OckeloenPRL2013, MuesselPRL2014}. Extending this process to 1D systems, the criticality of the TnT scheme is the underlying mechanism of the miscible-immiscible quantum phase transition at zero temperature \cite{SabbatiniPRL2011}. Such a system is ideally suited for studies of spatial quantum correlations and scaling behavior in pattern formation dynamics.

\begin{acknowledgments}
We thank Vladan Vuleti{\'c} for insightful discussions at the very beginning of the project. This work was supported by the Heidelberg Center for Quantum Dynamics and the European Comission FET-Proactive grant AQuS (Project No. 640800). W.M. acknowledges support by the Studienstiftung des deutschen Volkes. B. J-D. acknowledges financial support from the Generalitat de Catalunya Grant No. 2014SGR401 and support by the Ram\'on y Cajal program.
\end{acknowledgments}


\appendix

\section*{Appendix: Monte Carlo wave function simulation}
\label{SecMC}

\begin{figure}[b]
\includegraphics[width = 86 mm]{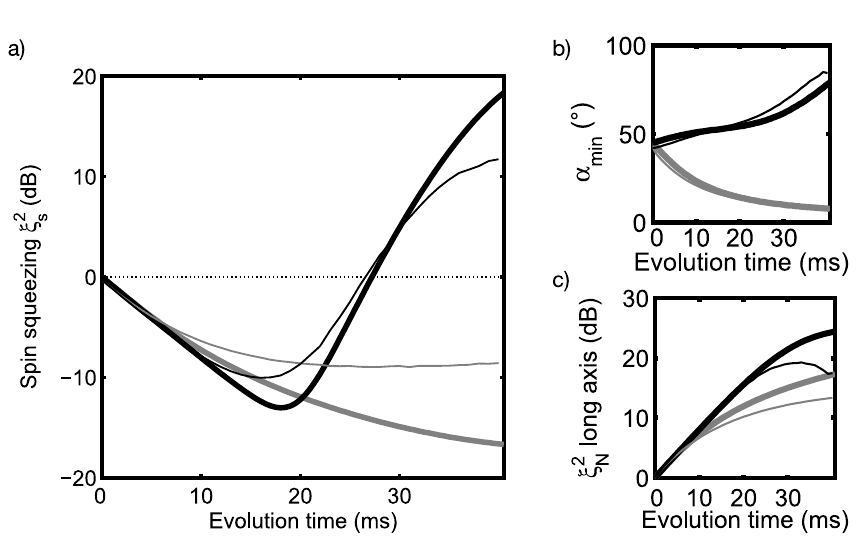}
\caption{{\bf Comparison of ideal theory and MCWF results.} The results of the MCWF simulations (thin black line for the twist-and-turn scheme and thin gray line for one-axis twisting) qualitatively agree with the ideal quantum evolution (thick solid lines) which does not include losses and the corresponding parameter changes for (a) spin squeezing parameter $\xi_{\text{S}}^2$, (b) optimal rotation angle $\alpha_{\text{min}}$, and (c) the fluctuations along the long axis of the state.  In both ideal theory and MCWF simulation, the spin squeezing rate of the twist-and-turn scheme exceeds the corresponding rate of one-axis twisting at intermediate evolution times, and squeezing is lost in the later evolution.}
\label{FigAppendixA}
\end{figure}

For the quantitative description of the experimental results (Fig.\,3 and 5), we perform numerical simulations using the Monte Carlo wave function method \cite{MolmerJOS1993,MolmerQSO1996, LiPRL2008}. This description includes the effects of atomic losses as well as the atom number dependencies of the parameters, which change due to varying atomic densities. For our calculations, we find good agreement using $\chi = 2\pi\times 1.43{\text{\,Hz}}/\sqrt{N}$, $\Omega = 2\pi\times 19$\,Hz and $\delta = - 2\pi\times 0.63(\sqrt{N}-\sqrt{550})$\,Hz. These parameters are consistent with independent measurements using plasma, $\pi$ as well as Rabi oscillations,  and the determination of collisional shifts from Ramsey sequences. Two-body spin-relaxation loss from $\upket$ and three-body Feshbach losses are included, with loss rates calibrated by independent measurements. 
The simulation also includes the spin-echo pulse at half the evolution time and detuning fluctuations with a standard deviation $\sigma_\delta =  2\pi\times0.45$\,Hz caused by variations of the bias magnetic field at 9.12\,G. These fluctuations are independently determined by repeated Ramsey measurements. All pulses are simulated with a Rabi frequency of $\Omega_{\text{pulse}} =  2\pi\times340$\,Hz and in the presence of the atomic nonlinearity, which leads to an effective shortening of the spin-echo rotation around the $x$ axis. \\
For each data point, we numerically calculate 8000 trajectories using a 4th order Runge-Kutta method and evaluate the observables by calculating the mean of the expectation values for the different trajectories. In each time step of a single trajectory, a random number determines if either the wave function is evolved according to an effective Hamiltonian  incorporating the ideal description Eq.\,1 with the addition of decay terms, or a loss event is implemented \cite{MolmerJOS1993,MolmerQSO1996, LiPRL2008}. This is done by properly cutting and renormalizing the evolved state vector and adjusting $\delta(N)$ and $\chi(N)$  accordingly. \\
In Fig.\,\ref{FigAppendixA}, the results of the MCWF simulation   are compared with the ideal theory. We assume an initial atom number of $N=500$ atoms and the time-averaged parameter $N\chi = 2\pi\times 30$\,Hz,  $\delta = 0$\,Hz and $\Omega =2\pi\times 19$\,Hz for the twist-and-turn scheme, and $\Omega = 0$\,Hz for one-axis twisting. On a qualitative level, we find good agreement between the ideal quantum evolution and the results of the MCWF simulations shown in Fig.\,3 of the main text. For the chosen parameters, the value for the optimal spin squeezing obtained from the ideal evolution is $\xi_{\text{S}}^2=-13.0$\,dB after an evolution time of $18$\,ms, while for the MCWF description we find optimal squeezing of $\xi_{\text{S}}^2=-10.1$\,dB after 16 ms.\\


\bibliographystyle{APS}


\end{document}